\documentclass[twocolumn,showpacs,showkeys,preprintnumbers,amsmath,amssymb]{revtex4}

\usepackage{graphicx}

\usepackage{dcolumn}

\usepackage{bm}

\begin{document}



\title{ Neutron-rich rare isotope production 
        from projectile fission of heavy beams \\
        in the energy range of 20 MeV/nucleon }


\author{N. Vonta$^{1,3}$}

\author{G.A. Souliotis$^{1}$}

\email[Corresponding author. Email: ]{soulioti@chem.uoa.gr}


\author{W. D. Loveland$^{2}$}

\author{Y. K. Kwon$^{3}$}

\author{K. Tshoo$^{3}$}

\author{S. C. Jeong$^{3}$}


\author{M. Veselsky$^{4}$}


\author{A. Bonasera$^{5,6}$}  

\author{A. Botvina$^{7,8}$}  


\affiliation{ $^{1}$ Laboratory of Physical Chemistry, Department of Chemistry, National and Kapodistrian
                     University of Athens, Athens 15771, Greece }

\affiliation{ $^{2}$ Department of Chemistry, Oregon State University, Corvallis, Oregon 97331, USA}

\affiliation{ $^{3}$ The Rare Isotope Science Project (RISP), Institute for Basic Science, 
                     Daejeon 305-811,   Korea.  }

\affiliation{ $^{4}$ Institute of Physics, Slovak Academy of
                     Sciences, Bratislava 84511, Slovakia        }

\affiliation{ $^{5}$ Cyclotron Institute, Texas A\&M University,
                     College Station, Texas 77843, USA           }

\affiliation{ $^{6}$ Laboratori Nazionali del Sud, INFN, via Santa Sofia 62, I-95123 Catania, Italy }

\affiliation{ $^{7}$ Frankfurt Institute for Advanced Studies, Goethe University, 
                     D-60438 Frankfurt am Main, Germany }

\affiliation{ $^{8}$ Institute for Nuclear Research, Russian Academy of Sciences, RU-117312 Moscow, Russia}


\date{\today}


\begin{abstract}

We investigate the possibilities of producing neutron-rich nuclides in projectile fission 
of heavy beams 
in the energy range of 20 MeV/nucleon  expected from low-energy facilities.
We report our efforts to theoretically describe the reaction mechanism of projectile fission 
following a multinucleon transfer collision  at this energy range. Our calculations are mainly
based on a two-step approach: the dynamical stage of the collision is  described with either 
the phenomenological  Deep-Inelastic Transfer model (DIT), or with the  microscopic Constrained Molecular 
Dynamics model (CoMD). 
The deexcitation/fission of the hot heavy projectile fragments is performed with the Statistical 
Multifragmentation Model (SMM). 
We compared  our model calculations with our previous experimental projectile-fission data of 
$^{238}$U (20 MeV/nucleon)+$^{208}$Pb and $^{197}$Au (20 MeV/nucleon)+$^{197}$Au
and found an overall reasonable agreement.
Our study suggests that projectile fission following peripheral heavy-ion collisions
at this energy range 
offers an effective route to access very neutron-rich rare isotopes 
toward and beyond the astrophysical r-process path. 

\end{abstract}


 \pacs{25.70.-z, 25.70.Hi,25.70.Lm}

 \keywords{ Heavy-ion nuclear reactions, projectile fission, peripheral collisions, deep-inelastic transfer, 
            constrained molecular dynamics, statistical multifragmentation, rare isotope beams, neutron drip-line,
            r-process
           }

\maketitle

\section{Introduction}

The study of the nuclear landscape toward the  astrophysical r-process path and the 
neutron drip-line is presently of special importance to the nuclear physics and astrophysics
community (see, e.g., \cite{ndrip00,ndrip01,Gade-2015} and references  therein). 
Closely related to this study  
is the efficient production of  very neutron-rich nuclides which is 
a central issue in current and future rare isotope beam facilities
(see, e.g., \cite{FRIB1,FRIB2,FRIB,GANIL,GSI,RIBF,ARGONNE,TAMU,EURISOL,EURISOL07,RISP,RISP-2013}).

Neutron-rich nuclides  are mainly produced  by spallation, fission and projectile  
fragmentation \cite{RIBreview2013}.
Spallation is an efficient mechanism to produce rare isotopes for ISOL-type techniques
\cite{Spallation,Spallation-2015}.
High-energy projectile fission is appropriate in the region of light and
heavy fission fragments (see, e.g., \cite{Ufission,Ufission12,Ufission13,Ufission15} for recent efforts on 
$^{238}$U projectile fission).
Finally, projectile fragmentation offers a universal approach to produce exotic nuclei 
at beam energies typically above 100 MeV/nucleon 
(see, e.g., \cite{MSUfrag1,MSUfrag2,MSUfrag3,Mazzocchi,Kurtukian,FragAdd1,FragAdd2}).
In addition to the above methods, peripheral reactions from the Coulomb barrier 
\cite{Volkov,Corradi,Watanabe,Loveland-2015,Wang-2016}
to the Fermi energy (20--40 MeV/nucleon) \cite{GS-PRC11} have been shown to produce very neutron-rich nuclei.
Such reactions offer the possibility of picking up a number of neutrons from
the target, in addition to the stripping of protons. We mention that our recent article \cite{Fountas-2014} 
elaborates on our current theoretical understanding and  computational description of the reaction mechanism 
for near projectile fragments for which substantial experimental effort has been
devoted by us in recent years \cite{GS-PRC11,GS-PLB02,GS-PRL03,GS-NIM03,GS-NIM08}.


Motivated by recent developments in several facilities that will offer intense heavy beams 
\cite{GANIL,ARGONNE,TAMU,RISP}  at the energy range 15--25 MeV/nucleon,
we wanted to understand the  possibilities of producing neutron-rich rare isotopes
employing projectile fission following  multinucleon transfer in this energy range. 
In this article, we will refer to this approach as quasiprojectile (QP) fission or, simply,  
projectile fission.
The mechanism  is essentially fission of a heavy quasiprojectile that results from 
extensive multinucleon transfer in the interaction  of the projectile with a heavy target. 
As it is the case for reactions with lighter non-fissionable projectiles \cite{GS-PRC11,Fountas-2014} 
in this energy range, the QP fission approach  offers the advantage of very broad  N/Z distributions
of the primary heavy QP that undergoes fission, thus leading to extremely (and possibly new) 
neutron-rich fission fragments. 
Moreover, the velocities of the projectile fission fragments can be high enough to allow 
adequate in-flight collection  and separation in appropriate large acceptance separators.


Projectile fission in the Fermi-energy range  was studied  by us some 
years ago with a 20 MeV/nucleon $^{238}$U beam interacting with a  $^{208}$Pb target
\cite{Souliotis-1997,Souliotis-1999}. 
In addition, projectile fission data of a 20 MeV/nucleon $^{197}$Au beam were also 
obtained as part of an extensive study of Au-induced reactions \cite{Souliotis-1998,Souliotis-2002}. 

In this paper, we present the experimental projectile fission data
of the 20 MeV/nucleon reactions $^{238}$U+$^{208}$Pb \cite{Souliotis-1997} and $^{197}$Au+$^{197}$Au
\cite{Souliotis-2002} and compare them with detailed calculations.
The dynamical calculations are  performed with either  the phenomenological  Deep-Inelastic Transfer 
model (DIT) \cite{DIT}, or with the  microscopic Constrained Molecular Dynamics model (CoMD) \cite{CoMD1}. 
The deexcitation (including fission) of the hot heavy projectile fragments is performed with the Statistical 
Multifragmentation Model (SMM) \cite{SMM,Botvina-2001}.
Furthermore, microscopic calculations with CoMD that follow the complete reaction dynamics
will also be presented following the developments reported in our recent article \cite{Vonta-2015}.

We will show that the present model framework is able to adequately describe the experimental data and, 
furthermore, may  allow dependable predictions of rates of neutron-rich rare  isotopes in various 
production and separation schemes employing the projectile fission approach.
This latter possibility is crucial for the selection and design of the proper separator configuration
and the optimum projectile/target combination (as well as, the appropriate target thickness) for efficient
production, collection and separation  of very neutron-rich nuclei from projectile fission.

The structure of the paper is as follows. 
In Section II, a short overview of our previous experimental measurements of projectile fission
at 20 MeV/nucleon is given. 
In Section III, a description of the theoretical model framework is presented.   
In Section IV, a systematic comparison of the 
production cross section calculations with the experimental data is given.  
Finally, a discussion and conclusions follow in Section V.

\section{Outline of the experimental methods and data}

The experimental data on 20 MeV/nucleon  $^{238}$U and $^{197}$Au projectile 
fission were obtained at the National Superconducting Cyclotron Laboratory 
(NSCL) of Michigan State University using the A1200 fragment separator \cite{A1200}.
The use of the A1200 for the  production and identification of medium and heavy mass
fragments has been discussed in detail in \cite{Souliotis-1998,Souliotis-2002}. 
For completeness, we briefly summarize the experimental setup and procedures.
The A1200 spectrometer was operated in the medium acceptance mode ͑with 
an angular acceptance $\Delta \Omega$ = 0.8 msr 
($\Delta \theta$ = 20 mr, $\Delta \phi$ = 40 mr) 
and a momentum acceptance $\Delta p$/$p$ = 3\%͒. In this mode, the A1200 
provided two intermediate dispersive images and a final achromatic image (focal plane).

For the U+Pb data, a 20 MeV/nucleon $^{238}$U$^{35+}$ beam 
with current 0.05 particle nA (3.1$\times$10$^{8}$ particles/sec) ͑struck a target of $^{208}$Pb ͑
(5.8 mg/cm$^2$ , 99.1\% enriched͒) at the object position of A1200 at an angle
of 1$^o$ relative to the optical axis of the spectrometer. 
Fission fragments from the decay of the projectile, arriving
at the first dispersive image of the spectrometer, passed
through a slit defining a 3\% momentum acceptance and an
X--Y position-sensitive parallel-plate avalanche counter
͑(PPAC) \cite{PPAC} ͑ giving a start timing signal͒. The horizontal position
of this PPAC, along with NMR measurements of the dipole 
magnetic field were used to determine the magnetic rigidity of the
particles.
At the focal plane, ͑14 m from the PPAC detector͒, the
fragments passed through a second PPAC ͑giving a stop timing signal͒
and then entered into a four-element ͑( 50, 50, 300 and 500 $\mu$m) Si 
detector telescope. 

For each event, time-of-flight, energy loss, residual energy  
and magnetic rigidity were recorded.
The spectrometer and the detectors were calibrated using a 
low intensity $^{238}$U beam and a series of analog beams.
From the measured quantities, the atomic number Z, the velocity $\upsilon$, the ionic charge q,
the mass number A, and the magnetic rigidity $B \rho$ were obtained for each event.
Further details of the procedure are reported in \cite{Souliotis-1998,Souliotis-2002}. 
Here we briefly mention that the determination of the atomic number Z 
was based on the energy loss $\Delta E_{1}$ and $\Delta E_{2}$ of the particles in the Si detectors
\cite{Hubert1,Hubert2}  and their velocity, with  resolution (FWHM) of 0.5 Z units. 
The ionic charge $q$ of the particles exiting the target  was obtained from
the total energy E$_{tot}$=$\Delta E_{1}$ + $\Delta E_{2}$ + E$_{r}$ (the last term being the 
energy in the stopping Si detector), the velocity $\upsilon$ and the magnetic rigidity   
via the magnetic rigidity equation  B$\rho $ = p/q (where p is the momentum).
The resulting ionic charge q had a resolution of 0.4 units (FWHM).
Since q must be an integer, we assigned integer values of q for each event 
by setting windows ($\Delta q=0.4$) on each peak of the q spectrum at each magnetic
rigidity setting of the spectrometer.
Using the  B$\rho $ and the velocity measurements, the mass-to-charge
A/q ratio  of each ion was calculated from the magnetic rigidity equation.
Finally, combining the q determination with the A/q measurement,
the mass number A was obtained as:
\begin{equation}
A = q_{int} \times A/q  \label{A_eqn}
\end{equation}
(q$_{int}$ is the integer ionic charge determined as described above) with an
overall resolution  (FWHM) of about 0.7 A units.   
To cover the mass and velocity range of the fission fragments, data were taken
at several overlapping  B$\rho $ settings of the spectrometer 
in the range 1.7--2.0 Tm.
These settings were such that only fission  fragments emitted forward in the rest 
frame of the quasiprojectile fissioning nucleus were accepted by the spectrometer. 
͑

To obtain production cross sections, the observed counts were normalized to the
beam current, target thickness and, furthermore, were corrected for the yields
of charge states missed by the focal plane detectors ͑due to charge changing of the ions
passing through the PPAC at the first image of the A1200͒. These  corrections
were based on the ionic charge distribution data of Leon et al.  \cite{Leon} and
were typically factors of 2--4.
Furthermore, the yields were corrected for the limited angular acceptance  of the 
spectrometer. For this correction, the angular distribution of the  fission fragments
was assumed to arise from an isotropic emission in the frame  of the fissioning nucleus
taken to be the projectile deflected on average  at an angle equal 
to the grazing angle $\theta_{gr} \sim$ 10$^{o}$ of the U+Pb reaction \cite{Wilcke}.

The Au+Au data were obtained under similar conditions as the U+Pb data, as part of our 
study of Au-induced reactions on heavy targets \cite{Souliotis-2002} which followed our
studies of Au+C,Al reactions \cite{Souliotis-1998}.
We note that the Au+Au data were only recently fully analyzed for the purpose of the present work
and shown in this article.
For these data,  a beam of $^{197}$Au$^{29+}$ (20 MeV/nucleon)
with current 0.5 particle nA (3.1$\times$10$^{9}$ particles/sec)
interacted with a 2 mg/cm$^{2}$ $^{197}$Au target and forward-moving fission
fragments were collected in the B$\rho $ range 1.4--1.8 Tm.
We point out that while a beam of $^{197}$Au (N/Z= 2.49) is not the optimum choice, 
compared to the $^{238}$U  (N/Z= 2.59) beam,  for the production of neutron-rich isotopes,  
the Au projectile fission data can serve as a complement of the U data for the detailed 
testing of our model framework, as will be discussed in the following.

\section{Brief description of the theoretical models}

The calculations on projectile fission performed in this work are based 
on the two-stage  approach typically employed for heavy-ion collisions. 
The dynamical stage of the collision was described with two models: 
the phenomenological Deep-Inelastic Transfer (DIT) model 
and the microscopic Constrained Molecular Dynamics (CoMD) model.
For peripheral and semiperipheral collisions, the dynamical stage of the
collision leads to an excited heavy projectilelike fragment 
(quasiprojectile) and an excited targetlike fragment (quasitarget).
In the second stage, the excited quasiprojectile was
deexcited with the Statistical Multifragmentation Model (SMM) in which
fission was properly taken into account.  
Furthermore, a complete dynamical description of the projectile--target interaction
was attempted with the CoMD model, along the lines of our recent work on 
proton-induced fission \cite{Vonta-2015}.
We briefly describe the above dynamical and statistical
codes and the calculation procedures in the  following.

\subsection{The Deep-Inelastic Transfer (DIT) model}

The  Deep-Inelastic Transfer (DIT) model of Tassan-Got 
\cite{DIT} is a phenomenological model that simulates stochastic nucleon exchange 
in peripheral and semiperipheral collisions.
The projectile and the target, assumed to be spherical, approach each other 
along Coulomb trajectories until they reach  the range of the nuclear interaction.
At this point the system is represented as two Fermi gases in contact.
A window opens in the inter-nuclear potential and nucleons are exchanged stochastically.
The direction and type of transfer are chosen by random drawing based 
on transfer probabilities. The transfer probabilities are calculated via a phase-space integral
that involves a phase-space flux term, the barrier penetrability and the occupation probabilities.
The nucleon transfer leads to  a variation in mass, charge, excitation energy and spin
of the interacting nuclei.
After interaction, the quasiprojectile and the quasitarget are excited and 
move along Coulomb trajectories. 
In the DIT model, the exchange of nucleons is assumed to be the only source of energy dissipation. 
Nucleon-nucleon collisions (mostly blocked by the Pauli principle
for low energy collisions) are not taken into account.
The calculations are performed for a wide range of impact parameters from very
peripheral to semiperipheral collisions.
The DIT model has been employed for the description of multinucleon transfer products from
our 15 and 25 MeV/nucleon $^{86}$Kr-induced reactions \cite{GS-PRC11,Fountas-2014,GS-PLB02,GS-PRL03}. 
Moreover  it has been successfully used in a number of other studies at Fermi energies 
(e.g. \cite{MV1,MV2,MV3,Keksis-2010,Souliotis-2014,Veselsky-2011}) and is able to describe correctly the N/Z, 
the excitation energy and kinematical properties of excited projectilelike (and targetlike)
residues emerging after the peripheral heavy-ion collision.


\subsection{The Constrained Molecular Dynamics (CoMD) model}

The  Constrained Molecular Dynamics (CoMD) model  
is a microscopic model  designed for reaction-dynamics studies near and below 
the Fermi energy \cite{CoMD1,CoMD2}. 
Following the approach of Quantum Molecular Dynamics (QMD) models \cite{QMD}, 
in the CoMD code, nucleons are described as localized Gaussian wave packets. The wave function 
of the nuclear system is assumed to be the product of these single-particle wave functions.
With this Gaussian description, the N-body time-dependent Sch$\ddot{o}$dinger equation leads to 
(classical) Hamilton's equations of  motion for the centroids of the nucleon wavepackets. 
The potential part of the Hamiltonian consists of a simplified Skyrme-like 
effective interaction and a surface term. The isoscalar part of the effective  interaction corresponds
to a nuclear matter compressibility of K=200 (soft EOS). 
For the isovector part, several forms of the density dependence of
the nucleon-nucleon symmetry potential are implemented in the code.
Two of them were used in the present work, that we named as
the standard potential and  the soft potential.
These forms correspond to a symmetry potential proportional to
the density and its square root,  respectively 
(see \cite{Papa-2013} and references therein).
We note that in the CoMD model, while not explicitly implementing
antisymmetrization of the N-body wavefunction, a constraint in the
phase space occupation for each nucleon is imposed, 
restoring the Pauli principle at each time step of the system 
evolution. This constraint restores in an approximate
way the fermionic nature of the nucleon motion in the interacting
nuclei.
The short range (repulsive) nucleon-nucleon interactions are
described as individual nucleon-nucleon collisions governed by the
nucleon-nucleon scattering cross section, the available phase space
and the Pauli principle, as usually implemented in transport codes.

In the present work, the CoMD code with its standard parameters
was used. 
The calculations were performed with either the standard  or the 
soft  symmetry potential and lead to nearly identical results for 
peripheral collisions.  Thus,  results with only the standard 
symmetry potential  will be presented in the following figures.
The ground state configurations of the projectile and the target nuclei
were obtained with a simulated annealing approach and were tested for stability
for relatively long times (1500--2000 fm/c). 
These configurations were used in the CoMD code for the 
subsequent collision simulations. 

As with the DIT model, the CoMD calculations for the dynamical stage
of the projectile--target interaction were performed in a wide range 
of impact parameters, following a triangular distribution, 
covering from peripheral to semiperipheral collisions.
The CoMD calculations were terminated at t=300 fm/c 
and the  characteristics of the produced excited
quasiprojectiles were determined with the fragment-recognition 
routine of CoMD.
This routine is based on the minimum spanning tree method 
assuming that nucleons with a relative distance of less than 2.4 fm belong to the
same fragment.
In the calculations, the excitation energy of the quasiprojectiles 
was obtained from the difference of their  binding energy  as given by the CoMD
calculation  at t=300 fm/c
and the corresponding binding energies of the ground-state nuclei taken 
from mass tables \cite{Moller}. 

\subsection{ Model for the de-excitation stage}

To describe the de-excitation of the hot heavy quasiprojectiles emerging from the 
dynamical stage, we used the latest version of the Statistical Multifragmentation Model    
\cite{SMM,Botvina-2001,Botvina-2005,Souliotis-2007,Botvina-2013}.
The SMM  model assumes statistical equilibrium at a low-density
freeze-out stage and includes all breakup channels ranging from the
compound nucleus to vaporization
(i.e. channels with only light particles A$<$4),
allowing  a unified description of nuclear disintegration
from low to high excitation energy.
In the microcanonical treatment, the statistical weight of a decay
channel is calculated as exponential of the entropy. Light fragments
with A$<$4 are considered as stable particles (nuclear gas)
with only translational degrees of freedom.
Fragments with A$>$4 are treated as heated liquid drops with free
energies parameterized as a sum of volume, surface, Coulomb and
symmetry energy terms \cite{Botvina-2001}
with parameters  adopted from the Bethe-Weizsacker mass formula.
The model generates a Markov chain of partitions (with the
Metropolis algorithm) representative of the whole partition ensemble.
In the low-density freeze-out configuration, the hot primary fragments are
assumed to be isolated and at normal density; they, subsequently, propagate in
their mutual Coulomb field, while undergoing binary deexcitation
via evaporation, fission or Fermi breakup \cite{SMM}.

The Coulomb interaction energy is directly calculated for each spatial
configuration of fragments in the freeze-out volume. 
Finally, the effect of the Coulomb field of the target
in proximity to the decaying quasiprojectile is included.
The evolution of the symmetry energy is taken into account 
in the mass calculation of primary and secondary fragments as in
\cite{Botvina-2005,Souliotis-2007}.
Below a threshold of excitation energy per nucleon  $\epsilon^{*}$=1 MeV, a smooth transition to standard 
experimental masses is assumed \cite{Botvina-2005}.
In the present calculations, the freeze-out density is  assumed to be 
1$/$3 of the normal nuclear matter density.

We note that for the deexcitation of low energy  ($\epsilon^*$ $<$ 1.0 MeV/nucleon) 
non-fissionable nuclei
(relevant for the production of neutron-rich isotopes \cite{Veselsky-2011,Fountas-2014}),
the SMM code has been shown to  adequately describe  the particle deexcitation process as 
a cascade of emissions of neutrons and light charged particles 
using the  Weisskopf-Ewing model of statistical evaporation.  
In regards to fission of heavier excited nuclei \cite{Botvina-2013},
the following approach is followed.
A "multifragmentation" threshold value of $\epsilon^*_{mult}$ = 0.2 MeV/nucleon
is defined,  above which the SMM statistical multipartition is applied, as described above.
This threshold value is, of course, much lower than the true nuclear multifragmentation threshold
of 2--3 MeV/nucleon, but it is employed in this work as a parameter to define when the 
SMM multipartition scheme will be  applied to the decay of the excited nuclear system.
Thus, for intermediate and high excitation energies ( $\epsilon^*$ $>$ $\epsilon^*_{mult}$ ),
fission is simply described as a special case of multifragmentation, i.e., 
the binary partition of the excited heavy nucleus.
However, at low excitation energy (  $\epsilon^*$ $<$ $\epsilon^*_{mult}$ ),
the fission channel is described in the spirit of the liquid-drop model
with deformation-dependent shell effects. 
In the present version of SMM, the Bohr-Wheeler approach \cite{Bohr-1939} is used 
for the calculation of the partial fission width.
The fission barrier is determined as in  Myers and Swiatecki \cite{Myers-1966}, 
and the results of ref. \cite{Il-1992} are used for the level density at the saddle point.
The method for obtaining fission mass distributions in the SMM  model
is described in detail in \cite{Botvina-2013}.
We briefly mention that 
along with a symmetric fission mode, two asymmetric fission modes are included 
with contributions  dependent on the fissioning nucleus and the excitation energy.
These contributions are described by an empirical parameterization based
on analysis of a large body of experimental data \cite{Botvina-2013}.
In closing, we point out that the current version of SMM, 
as described above, can adequately cover 
the conventional evaporation and fission processes
occurring at low excitation energy (i.e. compound-nucleus deexcitation),  
as well as the transition toward the higher energy multifragmentation
regime. In this regime, the binary (fission-like) partitions of the
excited heavy nuclei appear as a class of multifragmentation channels
competing with each other according to their statistical weight.

In the present work, apart from the value of
$\epsilon^*_{mult}$ = 0.2 MeV/nucleon that we used in 
the calculations presented in the following,
we also tested the values 0.5, 1.0 and 1.5 MeV/nucleon, 
and we  concluded that the choice of the low value of 0.2 MeV/nucleon 
is the optimum for the SMM description of the neutron-rich projectile 
fission fragments from the studied reactions.


\subsection{ Full CoMD description of quasiprojectile fission}

Apart from the two-step description using the approaches described 
above, we also employed the microscopic CoMD code in a stand-alone mode 
to perform calculations of the complete dynamics of the reaction
following the evolution for long times, up to t=15000 fm/c ( 5.0$\times$10$^{-20}$ sec ).
This CoMD calculation follows our procedures developed in \cite{Vonta-2015}
for the description of proton-induced fission of actinides.
For the present work, for a given reaction, a total of 1000 events were collected. 
For each event the impact parameter of the collision was chosen in the range b = 0--13 fm,
following a triangular distribution. 
The phase space coordinates were registered every 100 fm/c. 
At each time step, fragments were recognized and their properties were reported. 
From this information, we followed the evolution of the fissioning quasiprojectiles
and we obtained the properties of the resulting projectile fission fragments.
As in \cite{Vonta-2015}, we consider as fission time (t$_{fission}$) for the quasiprojectile, 
the time interval between  the projectile--target contact and the emergence of two
fission fragments from the quasiprojectile. Typical values of t$_{fission}$ are in
the range 2000--4000 fm/c.
Furthemore, we allowed an additional time t$_{decay}$ = 5000 fm/c after scission for the 
nascent projectile fission fragments to deexcite. 

\section{ Comparison of model calculations with data} 

In this section we present a systematic comparison of the calculations 
on projectile fission fragments 
from the 20 MeV/nucleon reactions $^{238}$U+$^{208}$Pb and $^{197}$Au+$^{197}$Au 
employing the model framework described above with our experimental data,
as discussed in Section II.

\subsection{Mass Yield and N/Z distributions}


In Fig. 1, the isobaric yield distributions (cross sections)  of projectile   
fission fragments from the reaction $^{238}$U (20 MeV/nucleon) + $^{208}$Pb
are presented. The measured cross sections (within the spectrometer acceptance)
are shown by the full (black) points. The open points (upper yield curve) are 
the total cross sections extracted after correction for the acceptance of 
the spectrometer \cite{Souliotis-1997}, as also explained in Sec.  II.
We note that the symmetric shape of the measured yield curve rules out any significant
contribution of the low-energy fission processes in the reaction mechanism (that,
as well known,  would lead to asymmetric fission due to shell effects).
The solid (red) lines are the results of the DIT/SMM calculations, whereas the  dotted (blue) 
lines are the CoMD/SMM calculations. The upper curves give the total cross sections. 
The lower curves are  obtained  after filtering of the corresponding calculations 
with the angular and B$\rho$ acceptance of the spectrometer.

From the figure, we see some agreement of the calculated isobaric yields with 
the corrected data,  mainly in the central region of the yield curve. The DIT/SMM
calculation appears to describe the peak of the mass yield distribution, but it is wider
than the measured distribution. 
We may interpret this by a possible incomplete collection of the heavier fission fragments
in the spectrometer due to B$\rho$ coverage. 
[We point out that collection of fission fragments was possible only when intense 
scattering of $^{238}$U primary beam at various charge states was not present.]

Furthermore, we compare the filtered calculated isobaric yields with  the measured data and 
we see an agreement again mainly in the central region of the yield curve. 
This provides confidence in our overall model framework regarding a realistic description 
of not only the yields  of the fission fragments, but also their kinematical properties and the
ionic charge state distributions. In our calculations, the ionic charge states of the fragments
have been simulated in a Monte Carlo fashion following the Gaussian parameterizations 
given in \cite{Leon}.  

Finally, the dashed (green) line is a complete CoMD calculation (1000 events, t=15000 fm/c) 
of projectile fission fragments and is in reasonable agreement with the two-step calculations 
as well as the data. We note that due to the relatively low statistics of this calculation, 
filtering  of the events with the spectrometer acceptance was not performed (thus, a corresponding 
filtered mass yield curve is not shown in Fig. 1).


In Fig. 2. we present the average Z/A (proton fraction) versus A of 
projectile fission fragments  from the reaction 
$^{238}$U (20 MeV/nucleon) + $^{208}$Pb.
The experimental data are given by the closed points \cite{Souliotis-1997}. 
The full (red) lines are calculations with DIT/SMM. The dotted (blue) lines
are calculations with CoMD/SMM. The continuous (black) line gives the line 
of beta stability.
The upper set of calculated curves are the final (cold) fragments 
produced after deexcitation of the primary (hot) fragments obtained at the 
partition stage of SMM. The average Z/A of these hot fragments is given by the 
lower set of curves. 
We observe that the DIT/SMM and CoMD/SMM calculations for the fission fragments
are on average less neutron-rich that experimental data. 
As we will also discuss in the following regarding Figs 3 and 4, we understand 
that the experimental data did not cover a low enough 
B$\rho$ range, so as to adequately collect the neutron-deficient fragments
that are products of high-energy fission processes.

Furthermore, the dashed (green) line is the full CoMD calculation with 
t$_{decay}$ = 5000 fm/c (see Sect. III-D). This line closely follows  
the two-step calculations CoMD/SMM and DIT/SMM for the hot fragments.
In other words,  the complete CoMD calculations result in fission fragments that 
are nearly as  neutron-rich as the hot fragments from the partition stage of the SMM 
calculation. 
Thus, these fragments are still  excited and require substantially  longer time
to evolve and give off the remaining of their excitation energy. Nonetheless, 
these observations give further  support of the two-stage approach we adopted 
to describe the present reactions.

Alond with our 20 MeV/nucleon U+Pb data and calculations, we show in Fig. 2 
the data of 750 MeV/nucleon U+Pb from \cite{GSI-1,GSI-2,GSI-3} 
obtained with the FRS separator at GSI. 
(We will call these data as ``GSI data'' in the following discussion.)
As described in the above references, the kinematical analysis of these data
allowed separation of the observed fragments in three groups 
according to their production mechanism: a) low-energy fission (dominated by Coulomb fission)
presented in Fig. 2 with right triagles, b) high-energy fission shown by squares
and c) fragmentation (abrasion-ablation) shown by inverted triangles.
From Fig. 2, we see that the low-energy fission GSI data are, as expected, very neutron-rich
and are close to our hot DIT/SMM (and CoMD/SMM) calculations for the 20 MeV/nucleon
U+Pb reaction.  The high-energy fission GSI data appear to follow our calculations for
the cold fragments.

Furthermore, the fragmentation-residue GSI data are very neutron-deficient. 
They are more neutron-deficient than our projectile fission data of $^{238}$U(20MeV/nucleon)+$^{27}$Al
\cite{Souliotis-1997,Souliotis-1999} that we present in Fig. 2 by open circles. 
We briefly mention that the 20 MeV/nucleon $^{238}$U + $^{27}$Al data were obtained with the A1200
spectrometer under the same conditions as the U+Pb data in the magnetic rigidity range of 1.60--1.80
Tesla-meters. This condition selected neutron-deficient fission fragments coming mostly from fusion-like events.
(Nuclide cross sections of these fragments are shown in Fig. 3 by open circles.)
Finally, in Fig. 2, the dashed  line shows the location of 
the radiochemical low-energy fission-fragment data of p(24 MeV) + $^{238}$U 
\cite{Kudo-1998}, which are, on average, more neutron-rich than our U+Pb
fission-fragment data and final (cold) calculations, but less neutron-rich 
than the GSI low-energy data.


\subsection{Mass distributions of selected isotopes}

In Fig. 3, we  first present the experimental mass distributions (closed points)
of several elements 
from  the projectile fission of $^{238}$U (20 MeV/nucleon) interacting with the
$^{208}$Pb target \cite{Souliotis-1997}.
The DIT/SMM calculations are given by the solid (red) lines and the CoMD/SMM 
calculations by the dotted (blue) lines. These two calculations are close to each other
on the neutron-rich side, while the latter calculations are systematically 
lower in the central region of the distributions (as we also observed in Fig. 1).
From the figure,  we observe some agreement in the shape of the calculations with the 
experimental cross sections. More specifically, we are able to obtain a relatively good 
description of the neutron-rich sides of the isotope distributions, which are of 
primary interest for the present work.
For the neutron deficient sides, the present calculations extend more to the left compared to the
experimental data. We explored this issue in detail and concluded that in the experimental data, the 
magnetic rigidity range did not extend to the low-B$\rho$ region necessary for these 
isotopes, as we mentioned before in relation to Fig. 2.
From a practical standpoint, it is clear that, in order to avoid such issues in upcoming experiments, 
careful consideration  should be given to the necessary of B$\rho$ coverage that is possible to be 
predicted with the help of the present model framework.
We note that along with the U+Pb calculations, we performed calculations
for projectile fission fragments from   (20 MeV/nucleon) $^{238}$U + $^{64}$Ni.
The fission-fragment cross sections are very close to those of the U+Pb reaction
(and are, thus, not shown in Fig. 3), but the angular  distributions are narrower 
(as we discuss later in relation to Fig. 5) rendering the use of the 
lighter target preferable for applications to neutron-rich RIB production. 
Moreover, in Fig. 3 the dashed  green lines (for Z = 54, 40) show the distributions of the hot fission
fragments from the DIT/SMM calculation (after the SMM partition, but before deexcitation). 
As we discussed before, these  products are very neutron-rich and deexcite toward more stable nuclides.

In Fig. 3, we also present the fission fragment data of $^{238}$U (20 MeV/nucleon) + $^{27}$Al
(open points) which, as we discussed in relation to Fig. 2, are comprised of neutron-deficient
nuclides. These data appear to complement the left side of the U+Pb data. They also seem to follow 
closely the left side of the DIT/SMM and CoMD/SMM calculations, implying their origin in
high-energy fission processes.
Interestingly, the nuclides on the left sides of the $^{238}$U + $^{27}$Al fission-fragment 
distributions  extend to very proton-rich nuclei toward  the proton-drip line, thus rendering 
such a reaction a possible pathway to access proton-rich nuclides.  
Detailed calculations for this reaction, comparison with the data,  as well as possible consideration
for proton-rich RIB production are currently underway.
 
Finally, in Fig. 3 we present the GSI data for $^{238}$U (750 MeV/nucleon) + $^{208}$Pb
as follows: a) the fission fragment data  coming from both low-energy and high-energy fission
are shown by right triagles and b) the fragmentation (abrasion-ablation) residue data are shown by
inverted triangles and appear to follow the left side of our U+Al fission fragment data.
As discussed in detail in \cite{GSI-3}, the fission-fragment cross sections of the GSI data for the neutron-rich 
nuclides in the region of asymmetric low-energy fission are especially large due to the prevalence of
electromagnetic fission (Coulomb fission). These cross sections (as shown for Z= 54, 50, 40, 36 in Fig. 3)
are larger than the 20 MeV/nucleon U+Pb data on the right side of the distributions.
However, in the region of symmetric mass division they are lower and are complarable to our data for U+Pb, as seen
for Z= 46, 42 in Fig. 3.
Nevertheless, it is interesting to note that the nuclide cross sections of our U+Pb data and the GSI data are comparable 
near the end of the neutron-rich sides in the full range of the measured fission fragments.
At this point we emphasize that the production of the most neutron-rich fragments in the 
relativistic  U+Pb reaction comes from low-E$^*$ (mostly Coulomb-induced) fission of the  $^{238}$U
projectile, whereas in the 20 MeV/nucleon U+Pb reaction, it comes from fission of low-E$^*$ U-like
quasiprojectiles after nucleon exchange with the target. Thus, the lower-energy regime of
20 MeV/nucleon (or lower) may offer the opportunity to go further in neutron-richness by exploiting the fission of
quasiprojectiles at the high N/Z and low E$^*$ tails of their distributions.
Consequently, projectile fission at energies below the Fermi energy can be a competitive means for
very neutron-rich RIB production at low energy facilities offering intense $^{238}$U beams.


In Fig. 4, we turn our attention to the projectile fission of $^{197}$Au (20 MeV/nucleon) 
interacting with a$^{197}$Au target. 
As we already mentioned, this reaction was studied  
as part of our extended study of Au-induced reactions at NSCL/MSU \cite{Souliotis-1998,Souliotis-2002}.
The Au+Au data were recently fully analyzed for the purpose of the present work.
The general behavior of the data in comparison to our calculations is similar to that 
of U+Pb (as discussed in Figs 1 and 2), we thus show here only the yield distributions of 
several fission fragments.
In Fig. 4, the  experimental mass distributions are shown by full points. 
We see some disagreement between the data and the DIT/SMM calculations 
[full (red) lines] and the CoMD/SMM  calculations [dotted (blue) lines] 
at the neutron-rich side  for the heavier fission fragments (Z=54, 50 in the figure).
We attribute these discrepancies, in part, to incomplete B$\rho$ coverage
during the experimental  measurements \cite{Souliotis-2002}.
For the lower elements, we observe that the CoMD/SMM calculations are in
good agreement with the data, whereas the DIT/SMM calculations are higher
than the data.
The differences between the calculations may be due to 
differences in the excitation energy distributions of the primary quasiprojectiles 
predicted by the two different dynamical codes.
Finally, in Fig. 4 the dashed  green lines (for Z = 54, 40) show the distributions of the hot fission
fragments from the DIT/SMM calculation (after the SMM partition, but before deexcitation). 
As also seen in Fig. 3 for U+Pb, these fragments are very neutron-rich and deexcite 
toward less exotic nuclides.
In closing, we note that the beam of $^{197}$Au (N/Z= 2.49) is not the optimum choice
for the production of neutron-rich isotopes in projectile fission, 
compared, of course,  to the $^{238}$U (N/Z= 2.59) beam. Nevertheless,
the Au+Au data offer additional detailed testing of our model framework.


\subsection{Angular distributions }

To understand the kinematics and the angular spread of the fission fragments 
from the projectile fission  of $^{238}$U at 20 MeV/nucleon, 
in Fig. 5a we show  the DIT/SMM calculated mass-resolved angular distributions for U+Pb.
The successive contours (starting from the innermost) represent  a drop in the yield
by a factor of two. 
The (lower) horizontal full lines represent the polar angular acceptance of the 
A1200 spectrometer in the experimental setup of \cite{Souliotis-1997} (Sect. II).
The (upper) horizontal dashed lines indicate the angular acceptance of the KOBRA separator
\cite{KOBRA-2015}, that we consider as a representative large acceptance separator
appropriate for rare isotope production at this energy regime. 
In the KOBRA setup, we assume that the beam hits the primary target at an angle of 
6$^{o}$ (that may be achieved with the aid of an appropriate beam swinger system
\cite{KOBRA-2015}) and fragments are collected in the polar angular range of 
8$^{o}$--18$^{o}$.
From the figure,  we visually comprehend the issue of the very small acceptance of our
setup in the original A1200 separator scheme. A fraction on the order of 1\% or smaller of the 
produced projectile fission fragments falls in the angular acceptance of the A1200
spectrometer, as is also indicated quantitatively in Fig. 1.
However,  with an advanced large-acceptance separator like KOBRA, a substantial fraction of
the projectile-fission fragments (that can reach 30--50\%) can be collected, 
provided that we swing the primary beam at an  appropriate angle 
(that we can choose to be near the grazing angle of the reaction).

In Fig. 5b, we show the mass resolved angular distributions for the reaction
of $^{238}$U (20 MeV/nucleon) with the ligher target of $^{64}$Ni. 
The grazing angle for this system is $\sim$4.0$^{o}$, much smaller than
the U+Pb system at 20 MeV/nucleon. 
We clearly see  that the angular distributions of the projectile fission fragments
are narrower, allowing, of course, a more efficient collection by the separator.
For this reaction, we may assume that the beam hits the primary target at an angle of 
3$^{o}$ in the KOBRA setup and fragments are collected in the polar angular range of 
5$^{o}$--10$^{o}$.

We note that our original choice of  $^{208}$Pb (N/Z=1.54) as a target \cite{Souliotis-1997} 
was based on its large N/Z.
However, with our present reaction model framework, we find that the calculated production 
cross sections of neutron-rich projectile fission fragments using the ligher $^{64}$Ni (N/Z=1.29) 
target  are nealy similar to those coming from reaction using the more neutron-rich $^{208}$Pb target. 
As we discussed in the previous section,
this is mainly due to the fact that the most neutron-rich fission fragments come from rather 
cold quasiprojectiles that, in turn, originate from  peripheral collisions of
the $^{238}$U projectile with the targets. 
Because of this, we may expect only a small effect in the  absolute cross sections  that we wish to 
investigate in detail both theoretically and experimentally. 
However, from a practical standpoint, the use of lighter targets is preferable, as it leads to narrower
angular distributions and thus larger acceptance which is desirable for applications of projectile fission
in RIB production schemes.
We mention that our event-by-event simulations may allow full event tracking of the products 
through the beam-optical elements of the separator. We can thus determine rates for 
the production and separation of desired neutron-rich projectile fission fragments,
as it is now being performed by members of the KOBRA team \cite{KOBRA-2015}.


\subsection{Production cross sections and rates of accessible neutron-rich nuclides}

After the above discussion on production cross sections and angular distributions, 
we wish to  provide an overall perspective of accessible neutron-rich nuclides in 
$^{238}$U projectile fission at 20 MeV/nucleon.
In Fig. 6b we show the DIT/SMM calculated production cross sections of 
projectile fission fragments  from  the reaction $^{238}$U (20 MeV/nucleon) + $^{64}$Ni 
on the Z--N plane.
Several cross section ranges are shown by open circles according to the figure key.
The closed squares show the stable isotopes. The solid (red) line shows the
astrophysical r-process path and the dashed (grey) line indicates the location
of the neutron drip-line as calculated by \cite{Moller}.
We observe that a broad range of very neutron-rich nuclides above Fe (Z=26)
becomes accessible with the projectile fission approach at this energy range.
Apart from the observable (cold) fission fragments (Fig. 6b), we show in Fig. 6a 
the hot fission  fragments  produced at the partition stage of SMM.
As expected, these hot fragments are extremely neutron-rich and extend
toward the r-process path and the neutron dripline. 
These isotopes are, in principle,  not accessible experimentally. However, 
some of their properties may be inferred from detailed calculations
(see e.g. \cite{Botvina-2015} and references therein). 
Furthermore, from an experimental point of view, kinematical reconstruction
of these hot  products may be possible taking advantage of modern 
neutron and/or parcticle multidetectors 
(see e.g. \cite{Wada-2014} and references therein).
 


From a practical standpoint,
it is interesting to evaluate, using the above  cross section calculations, what 
total rates are expected for very short-lived neutron-rich nuclei.
For this purpose, we assume a primary $^{238}$U beam intensity of 10 particle nA
(6.2$\times$10$^{10}$ particles/sec) at 20 MeV/nucleon from, e.g,  
the RISP accelerator complex \cite{RISP}.
Furthermore, we assume a  production target thickness of   20\,mg/cm$^{2}$ $^{64}$Ni,  
a separator angular acceptence  of 20 msr and momentum  acceptance of 5\%.
Under these conditions, we estimate that a production cross section of 
10 $\mu$b  corresponds to a production rate  of about 5 counts/sec. 
Of course, the overall transmission of the separator has to be carefully taken 
into account along with the above  estimate of total rates.
However, we may conclude that for a large number of very neutron-rich nuclei,
production rates of 10-1000 /sec (at energies around 20 MeV/nucleon) are possible, 
allowing  the study of the structure of these nuclei.

Furthermore, we point out  that for very neutron-rich nuclei toward the r-process path,
the predicted halflives are less than 1 second,  making their production 
in an ISOL facility very difficult.
For such nuclei, counting rates of 10--1000 /day should be reachable with 
the present projectile fission approach,  which  suffices to verify their stability
and in the most  favorable cases allow measurements of their decay properties.
Finally, an interesting observation from Fig. 6, is the possibility to 
move close to (or even reach) the neutron drip-line in the region  Z=45--50 
(A=130--140) with the present projectile-fission approach.


\subsection{Discussion and Plans}

We would like to conclude with some comments on the model approaches 
used in this work.
Starting from the full microscopic CoMD approach, we think that the overall 
successful description of the reaction is especially 
valuable due to the predictive power of the microscopic many-body approach, 
as we have also seen in our recent works \cite{Fountas-2014,Vonta-2015} that does not depend 
on ad hoc assumptions of the reaction dynamics.
However, the full CoMD description  is very computer intensive  and not practical, 
but, essentially validates the two-step approach as we saw in this work.
As a next step,  the two-stage CoMD/SMM approach provided very good results, 
but is still somewhat computer intensive (due to the CoMD stage of the calculation).
Finally, the phenomenological DIT/SMM approach, offering results nearly similar to the 
CoMD/SMM approach is fast and, thus, can be practical for the design of  experiments 
based on projectile fission  at this energy.

In the near future, apart from further calculational efforts,
we plan to perform detailed measurements of projectile fission of $^{238}$U 
at 10--15 MeV/nucleon  a) at Texas A\&M with the MARS recoil separator \cite{TAMU,GS-PRC11} 
and b) at LNS/INFN 
with the MAGNEX large-acceptance spectrometer \cite{MAGNEX}.
We expect that these measurements will provide  a detailed testing ground 
for our models, and will offer access to very neutron-rich nuclei for decay studies.
Furthermore, these efforts will provide experience and preparation for future plans
at upcoming large-acceptance separator facilities (e.g. KOBRA \cite{KOBRA-2015}).
In closing, we believe that the possibility of producing very neutron-rich nuclides 
from projectile fission at low energy facilities may allow a rich and diversified program
of nuclear structure studies of neutron-rich rare isotopes in these facilities,  
complementary to the current successful programs of higher-energy RIB facilities
\cite{RIBF,FRIB,GSI}.

\section{Conclusions}  

Summarizing, we investigated
the possibilities of producing neutron-rich nuclides in projectile fission 
of heavy beams 
in the energy range of 20 MeV/nucleon.  
We reported our efforts to theoretically describe the reaction mechanism of projectile fission 
following a peripheral collision  at this energy range. Our calculations are mainly
based on a two-step approach: the dynamical stage of the collision is  described with either 
the phenomenological  Deep-Inelastic Transfer model (DIT), or with the  microscopic Constrained Molecular 
Dynamics model (CoMD). 
The deexcitation/fission of the hot heavy projectile fragments was performed with the Statistical 
Multifragmentation Model (SMM) with appropriate settings of its parameters. 
We also employed the CoMD model for the full description of the reaction.
We compared  our model calculations with our previous experimental projectile-fission data of 
$^{238}$U (20 MeV/nucleon)+$^{208}$Pb and $^{197}$Au (20 MeV/nucleon)+$^{197}$Au
and found an overall reasonable agreement.
Our study suggests that projectile fission following peripheral heavy-ion collisions
at this energy range (i.e. well above the Coulomb barrier, but below the Fermi energy), 
offers an effective route to access very neutron-rich rare isotopes 
toward the astrophysical r-process path and possibly the neutron drip-line.  



\section{Ackowledgements}
\par
 
We are thankful to Dr. L. Tassan-Got for the DIT code and to 
Dr. M. Papa for his version of the CoMD code. 
We also thank Prof. Athena Pakou for numerous enlighting 
discussions and suggestions.
N.V. acknowledges fruitful discussions with Dr. Sato Yoshiteru.
Financial support for this work was provided, in part, by
the National and Kapodistrian University of Athens
under ELKE Research Account No 70/4/11395
and, in part, 
by the Rare Isotope Science Project of the Institute for Basic Science
funded by the Ministry of Science, ICT and Future Planning 
and National Research Foundation of Korea.
A. Botvina acknowledges the support of BMBF (Germany).
M.V. was  supported by the Slovak Scientific Grant Agency under contracts 2/0105/11
and 2/0121/14 and by the Slovak Research and Development Agency under contract
APVV-0177-11.


\bibliography{QPFission_study_20amev}             

\newpage



\begin{figure}[p]                                        
\centering
\includegraphics[width=0.45\textwidth,keepaspectratio=true]{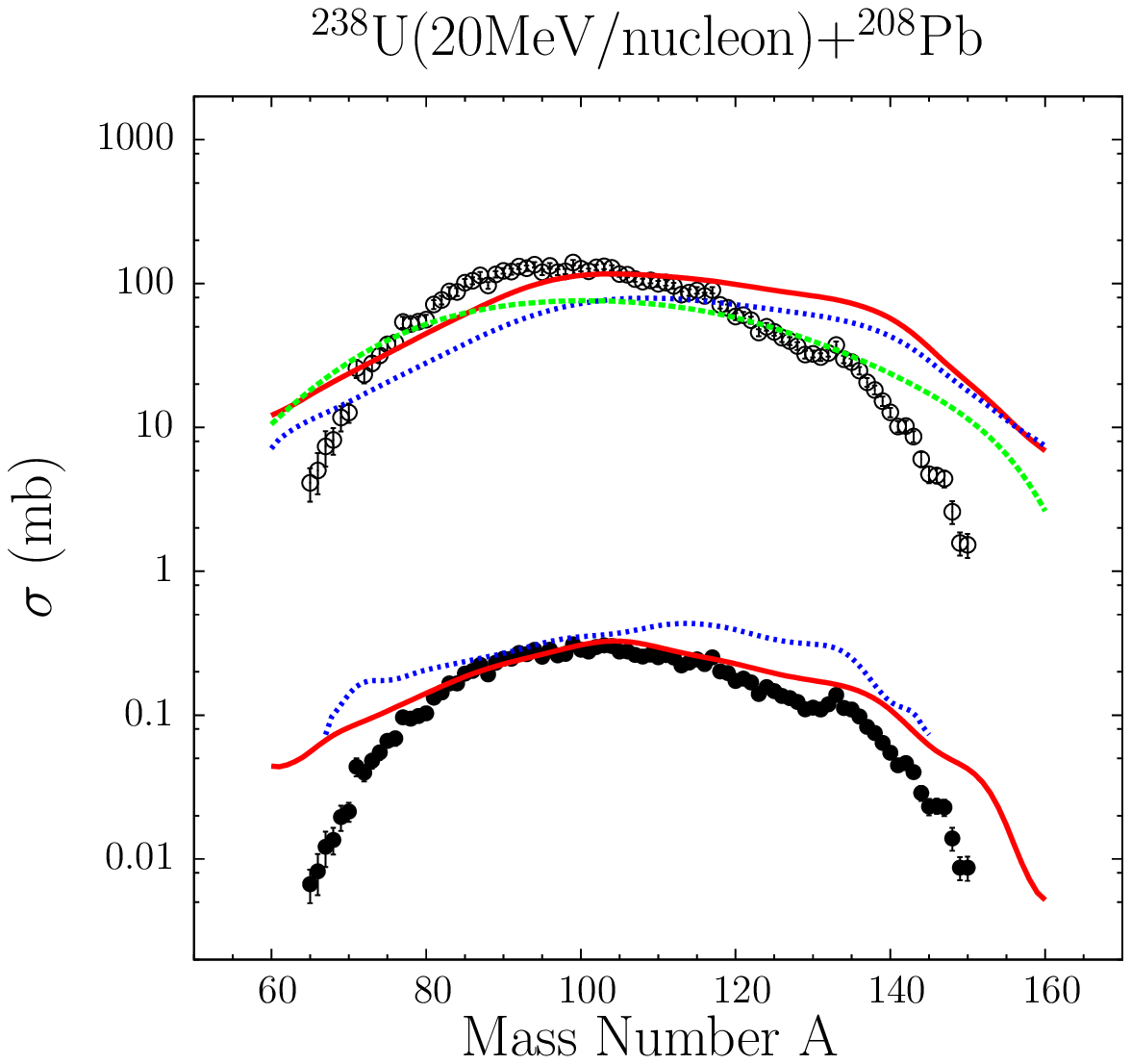}   
\caption{
(Color online) Isobaric yield distributions (cross sections)  of projectile   
fission fragments from the reaction $^{238}$U (20 MeV/nucleon) + $^{208}$Pb.
Closed points: measured cross sections, open points: total cross sections
extracted after correction for the angular acceptance of the spectrometer
\cite{Souliotis-1997}.
Solid (red) lines: DIT/SMM calculations. Dotted (blue) lines CoMD/SMM calculations.
The upper set of lines gives total cross sections. The lower set of lines is after filtering
with the acceptance of the spectrometer (see text).
The dashed (green) line gives the  total cross sections from the CoMD calculation of the 
full dynamics (1000 events, t=15000 fm/c).
}
\label{pf_fig01}
\end{figure}


\begin{figure}[h]                                        
\centering
\includegraphics[width=0.45\textwidth,keepaspectratio=true]{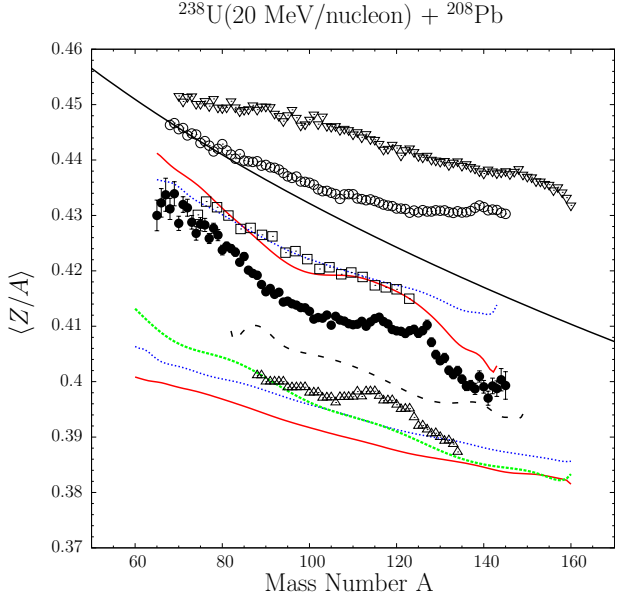}  
\caption{
(Color online) Average Z/A versus A of projectile fission fragments 
from the reaction $^{238}$U (20 MeV/nucleon) + $^{208}$Pb.
The experimental data for U+Pb are given by the closed points \cite{Souliotis-1997}.
The full (red) lines are calculations with DIT/SMM. The dotted (blue) lines
are calculations with CoMD/SMM.
The upper set of curves are the final (cold) fission fragments.
The lower set of curves are the hot fission fragments (after the SMM
binary partition, but before deexcitation).
The dashed (green) line is the full CoMD calculation (1000 events, t=15000 fm/c) with 
t$_{decay}$ = 5000 fm/c (see text). 
The open circles are data from $^{238}$U (20 MeV/nucleon) + $^{27}$Al \cite{Souliotis-1997,Souliotis-1999}.
The open triagles, squares and inverted triangles are the GSI data from the
reaction $^{238}$U (750 MeV/nucleon) + $^{208}$Pb for low-energy fission, 
high-energy fission and fragmentation products, respectively  \cite{GSI-1,GSI-2,GSI-3}.
The dashed (black) line represents the radiochemical data of p(24MeV) + $^{238}$U 
\cite{Kudo-1998}
The continuous (black) line gives the line of beta stability.
}
\label{pf_fig02}
\end{figure}


\begin{figure}[h]                                        
\centering
\includegraphics[width=0.45\textwidth,keepaspectratio=true]{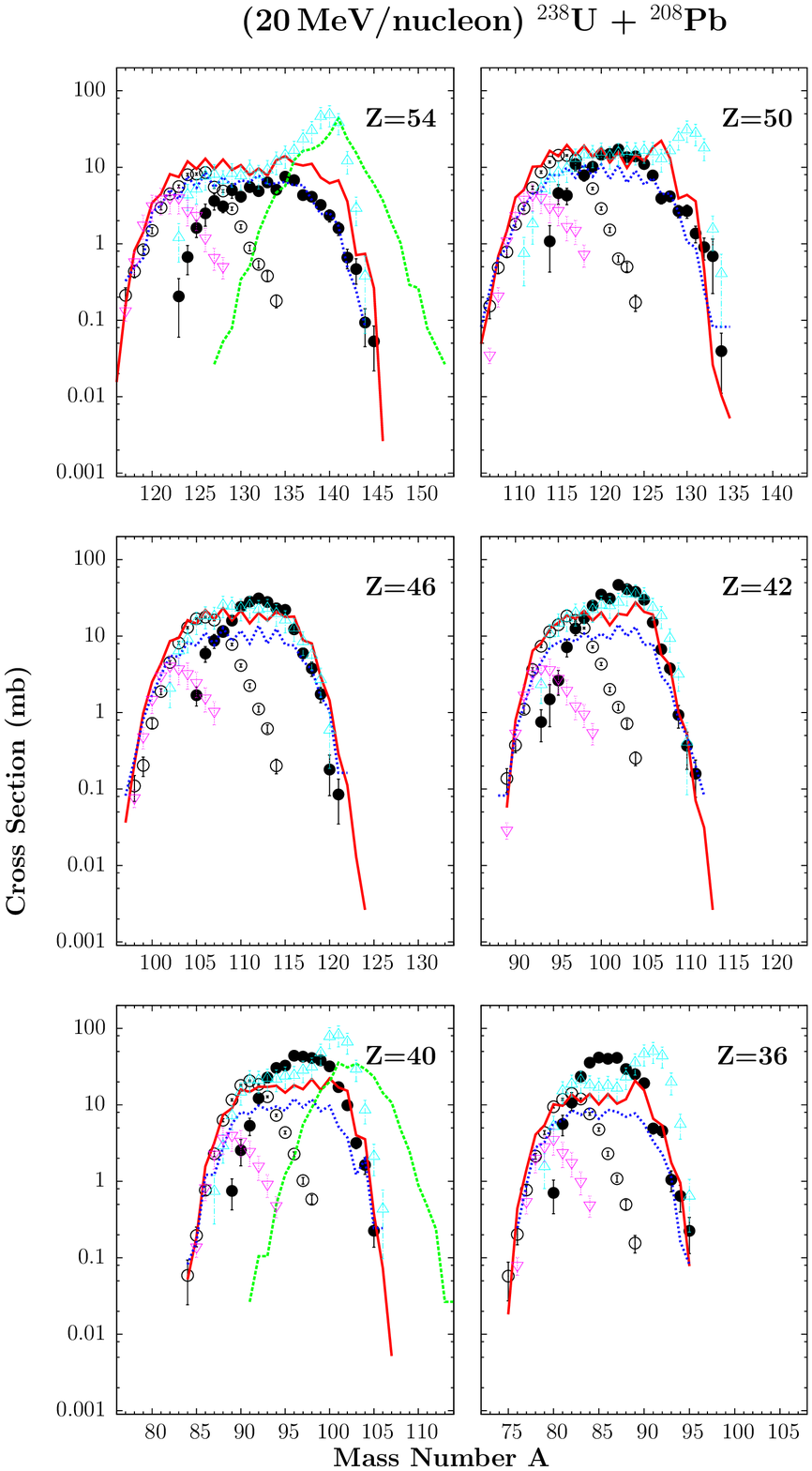} 
\caption{
(Color online) Comparison of calculated mass distributions (lines) of  
projectile fission fragments from the reaction $^{238}$U (20 MeV/nucleon) + $^{208}$Pb 
with the experimental data (closed points) of \cite{Souliotis-1997}. 
The calculations are with DIT/SMM [solid (red) line] and with CoMD/SMM [dotted (blue) line].
The dashed (green) lines for Z=54,40 are for hot fission fragments from DIT/SMM.
The open circles are data from $^{238}$U (20 MeV/nucleon) + $^{27}$Al 
\cite{Souliotis-1997,Souliotis-1999}.
The open right triagles and inverted triangles are the GSI data from the
reaction $^{238}$U (750 MeV/nucleon) + $^{208}$Pb for projectile fission and  
fragmentation (abrasion-ablation) products, respectively  \cite{GSI-1,GSI-2,GSI-3}.
}
\label{pf_fig03}
\end{figure}


\begin{figure}[h]                                        
\centering
\includegraphics[width=0.45\textwidth,keepaspectratio=true]{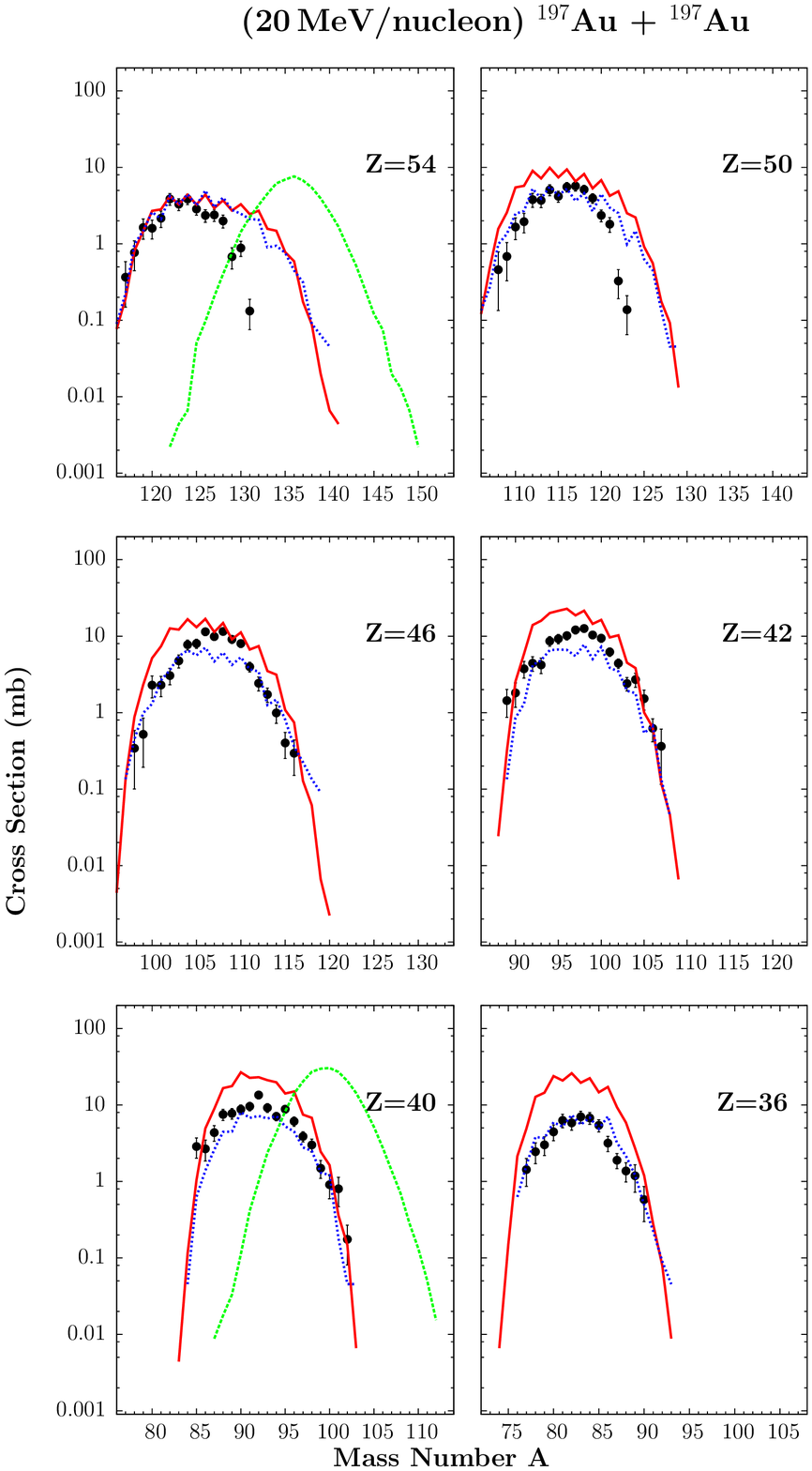} 
\caption{
(Color online) Comparison of calculated mass distributions (lines) of  
projectile fission fragments from the reaction $^{197}$Au (20 MeV/nucleon) + $^{197}$Au 
with the experimental data (closed points) of \cite{Souliotis-2002}. 
The calculations are with DIT/SMM [solid (red) line] and with CoMD/SMM [dotted (blue) line].
The dashed (green) lines for Z=54,40 are for hot fission fragments from DIT/SMM.
}
\label{pf_fig04}
\end{figure}

\begin{figure}[h]                                        
\centering
\includegraphics[width=0.45\textwidth,keepaspectratio=true]{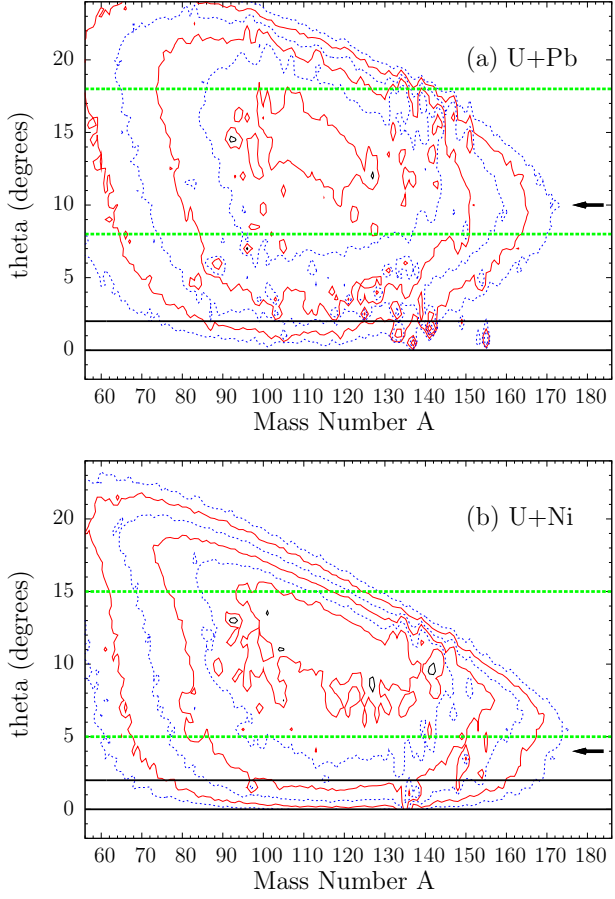}  
\caption{
(Color online) (a) DIT/SMM calculated mass-resolved angular distibutions of projectile
fission fragments from the reaction  $^{238}$U (20 MeV/nucleon) + $^{208}$Pb.  
The successive contours (starting from the innermost) represent  a drop in the yield
by a factor of two. 
The horizontal lines represent the polar angular acceptance of:
the A1200 spectrometer setup \cite{Souliotis-1997} (lower solid lines) and the KOBRA spectrometer
\cite{KOBRA-2015} (upper dashed lines)  (see text).
The arrow indicates the grazing angle of the U+Pb reaction (in the lab system).
(b) As in (a), but for the reaction  $^{238}$U (20 MeV/nucleon) + $^{64}$Ni (see text).
}
\label{pf_fig05}
\end{figure}


\begin{figure}[h]                                        
\centering
\includegraphics[width=0.45\textwidth,keepaspectratio=true]{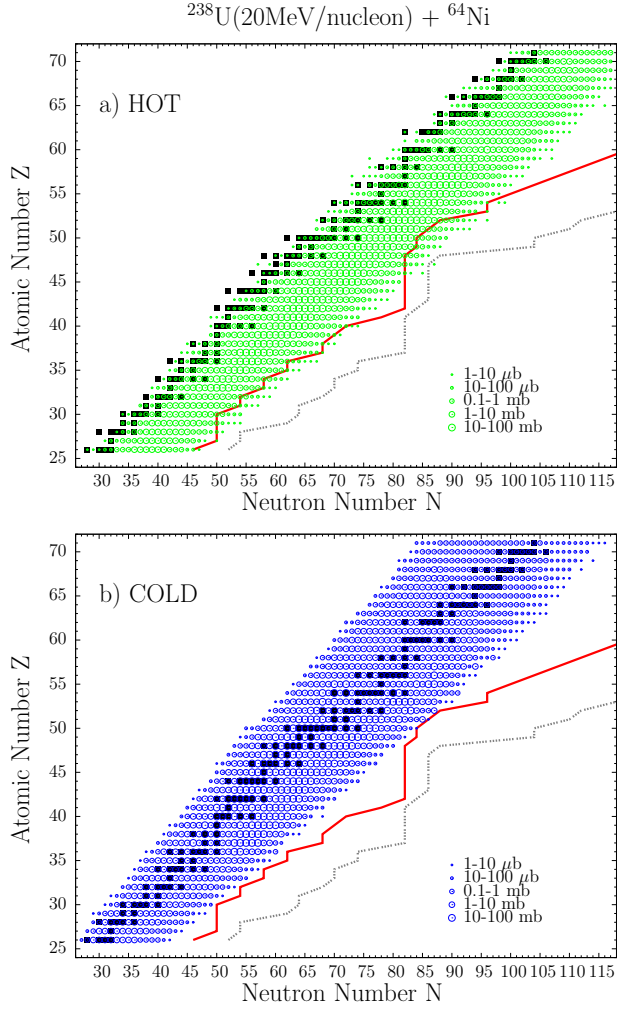}  
\caption{
Representation of DIT/SMM calculated  cross sections of projectile fission fragments from
          the reaction $^{238}$U (20 MeV/nucleon) + $^{64}$Ni on the Z--N plane.
          The cross section ranges are shown by open circles according to the key.
          Panel (a) shows the hot fission fragments (before deexcitation). Panel (b) shows 
          the final (cold) fission fragments.
          The closed squares are the stable isotopes.
          The solid (red) line shows the
          astrophysical r-process path and the dotted (grey) line shows  the location
          of the neutron drip-line  according to \cite{Moller}.
}
\label{pf_fig06}
\end{figure}


\end{document}